\documentclass[twocolumn]{elsart}

\usepackage{epsf}

\usepackage{natbib}

\usepackage{amssymb}

\begin{document}

\begin{frontmatter}



\title{The physics of jets in FR\,I radio galaxies}


\author[label1,label2]{R.A. Laing},
\author[label1]{J.R. Canvin},
\author[label3]{A.H. Bridle}
\thanks[AUI]{The National Radio
Astronomy Observatory is a facility of the National Science Foundation
operated under cooperative agreement by Associated Universities, Inc.}

\address[label1]{Astrophysics, University of Oxford, Denys Wilkinson Building, Keble
  Road, Oxford OX1 3RH, U.K.}
\address[label2]{Space Science and Technology Department, CLRC
  Rutherford Appleton Laboratory, Chilton, Didcot, Oxon. OX11 0QX, U.K.}
\address[label3]{National Radio Astronomy Observatory, 520 Edgemont Road, 
     Charlottesville, VA 22903-2475, U.S.A.}

\begin{abstract}
We model jets in low-luminosity (FR\,I) radio galaxies as intrinsically
symmetrical, axisymmetric, decelerating relativistic flows with transverse
velocity gradients. This allows us to derive velocity fields and the
three-dimensional distributions of magnetic-field ordering and rest-frame
emissivity. A conservation-law analysis, combining the kinematic model
with X-ray observations of the surrounding IGM, gives the profiles of
internal density, pressure, Mach number and entrainment rate along the
jets. We summarize our recently-published results on 3C\,31 and outline
new work on other sources and adiabatic jet models.
\end{abstract}


\end{frontmatter}

\section{Introduction}
\label{Intro}

This paper is a progress report on a project whose aim is to derive
quantitative estimates for the physical parameters of jets. We have
developed techniques to derive the three-dimensional distributions of
velocity, rest-frame emissivity and magnetic-field structure and hence to
deduce the jet dynamics via a conservation-law approach.  Our fundamental
assumption is that jets may be approximated as intrinsically symmetrical,
time-stationary, axisymmetric relativistic flows. We model their observed
radio synchrotron emission in total intensity and linear polarization,
using the observed differences between approaching and receding jets to
constrain velocity, emissivity and field. We then combine this model of
jet kinematics with a description of the surrounding IGM and use
conservation of energy, momentum and particles to estimate the internal
pressure and density.  For this technique to work, we need sources with
two-sided but asymmetrical and straight radio jets. We then make deep
observations with many resolution elements to derive total intensity and
linear polarization, corrected for any Faraday rotation. We also require
X-ray observations of the hot gas surrounding the radio source from which
the external pressure and temperature may be derived. At present, these
requirements are met only by VLA and {\sl Chandra} observations of nearby
FR\,I radio galaxies.

\section{3C\,31}
\label{3C31}

\begin{figure*}
\epsfxsize=14cm
\epsffile{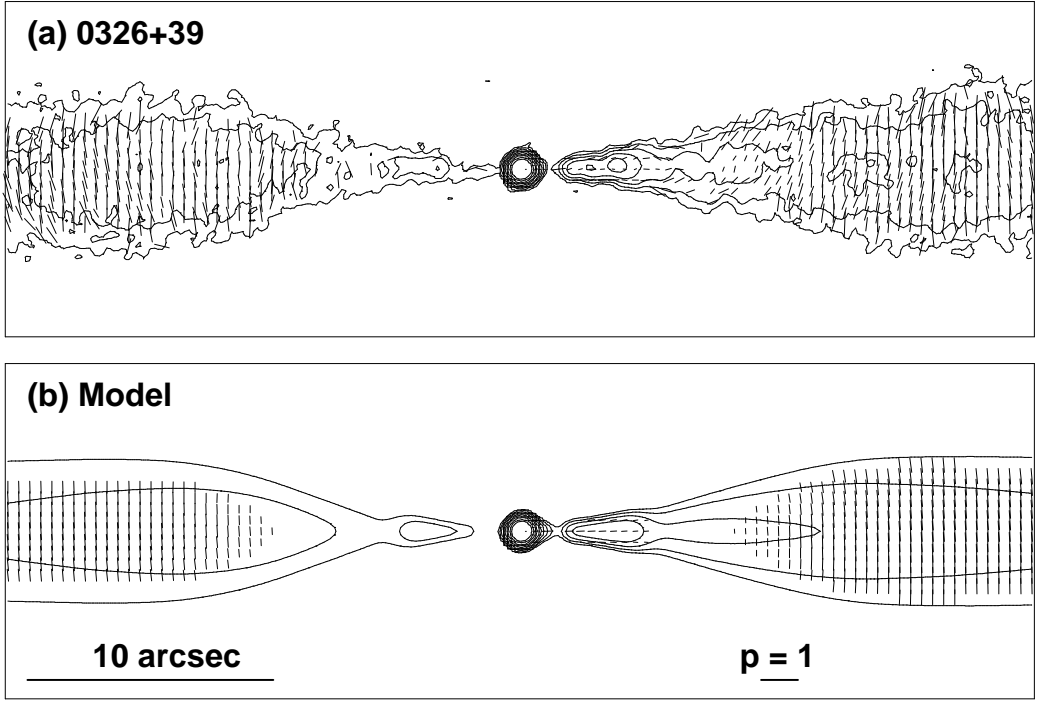}
\caption{Contours of total intensity superimposed on vectors with lengths proportional
  to the degree of polarization, $p$, and directions representing the apparent
  magnetic field. (a) VLA observations; (b) model. The resolution is 0.5\,arcsec
  FWHM and the contour levels are 1, 2, 4, 8, 16, \ldots $\times$ 25\,$\mu$Jy /
  beam area. The angular and polarization vector scales are shown by the
  labelled bars.
\label{0326fig}}
\end{figure*}

The first source to which we have applied these techniques in full is
3C\,31: the kinematic model, X-ray observations and conservation-law
analysis are described by \citet{LB02a}, \citet{Hard02} and \citet{LB02b},
respectively.  The kinematic model requires the jets to be at $\theta
\approx$ 52$^\circ$ to the line of sight.  The jets may be divided into three
distinct sections by geometry, velocity and emissivity variation: a
well-collimated inner region, a flaring region in which the jets widen
rapidly and then recollimate and a conical outer region.  Both
deceleration (from $\beta = v/c \approx 0.8$ where the jet flares to
$\beta \approx 0.25$ at the end of the modelled region) and transverse
velocity gradients are inferred. The toroidal magnetic field component is
larger than the longitudinal component almost everywhere and the radial
component is small except at the edge of the jet in the flaring region,
where the field is roughly isotropic, perhaps as a result of turbulent
entrainment.  The beginning of the flaring region marks a discontinuity in
emissivity and, probably, velocity.

The conservation-law analysis shows that the jets must be extremely light ($\rho
\approx 10^{-27}$\,kg\,m$^{-3}$). The entrainment rate has a local maximum where
the expansion is fastest; thereafter, it increases smoothly and
monotonically. The required entrainment could be provided by stellar mass loss
close to the nucleus, but interaction with the external medium is required at
larger distances. The jet is overpressured with respect to the surrounding
medium where it flares, but the outer region is likely to be in pressure
equilibrium.  The composition of the jet is not determined uniquely by this
analysis, but an electron-positron jet with entrained thermal matter would be
consistent with all of the available evidence.

\section{New results}
\label{B2}
 
We have applied simplified kinematic models to describe less detailed
observations of a complete sample of FR\,I radio galaxies from the B2
sample \citep{LPdRF}. Two of these, 0326+39 and 1553+24, have been
reobserved with the VLA at 8.4\,GHz and modelled in detail (Canvin \&
Laing, in preparation).  Good fits were again obtained: we show a
comparison between model and data for 0326+39 in Fig.~\ref{0326fig}.  The
basic picture of jet deceleration with tranverse velocity gradients holds
for both sources, but there are interesting differences in the field
structures: 1553+24 (like 3C\,31) has a dominant toroidal component but
the outer part of 0326+29  (Fig.~\ref{0326fig}) has roughly equal radial
and toroidal components, but no longitudinal field \citep[cf.][]{L80}.  

Thus far, we have assumed that emissivity and field structure are
independent of the velocity field. This is appropriate if we aim to deduce
the internal parameters without imposing preconceptions about the
underlying physics, but to make further progress, we need to make
additional assumptions.  The simplest approximation is that the jets are
``adiabatic'' in the sense defined by \citet{Bur79}: the relativistic
particles lose energy only by the adiabatic process and the magnetic field
is convected passively with the flow. \citet{Bau97} derived analytical
relations for the surface-brightness of a relativistic jet with no
transverse velocity gradient and either a transverse or longitudinal
magnetic field. We have generalized their approach to include shear and
arbitrary initial field geometry, using a formalism based on that of
\citet{MS}.

\begin{figure}
\epsfxsize=6.5cm
\epsffile{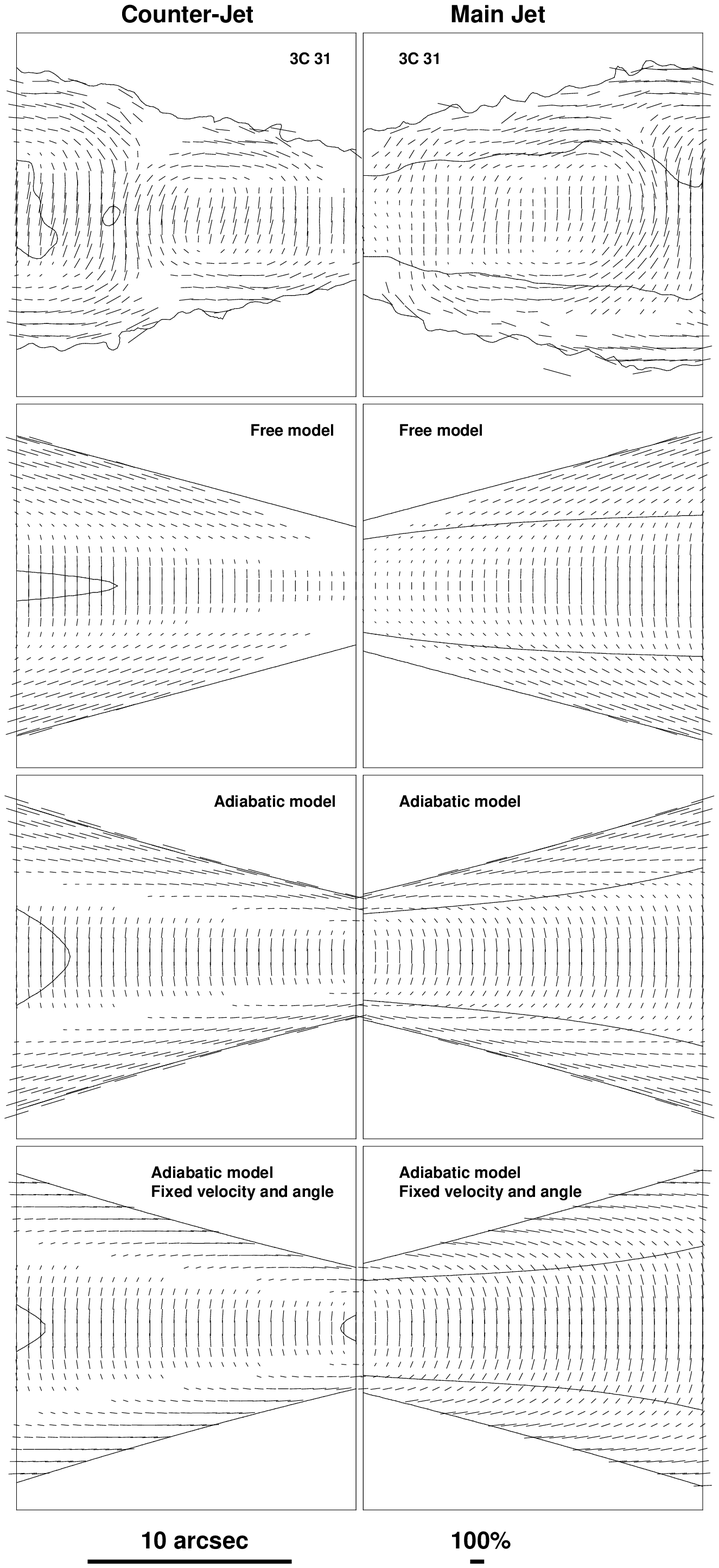}
\caption{Vectors with magnitudes proportional to the degree of polarization and
directions of the apparent magnetic field superimposed on 
contours of total intensity at a resolution of 0.75 arcsec for the outer regions
of the jets in 3C\,31.  The plots cover the range 10 --
27\,arcsec on either side of the nucleus.  Right: main jet; left: counter-jet.
From the top: observations; free Gaussian model from \citet{LB02a}; adiabatic
model with initial conditions set at the outer boundary and optimized velocity
and angle to the line of sight; adiabatic model with geometry and velocity from
free model. 
\label{3C31ad}}
\end{figure}

In Fig.~\ref{3C31ad}, we show the results of fitting adiabatic models to the
outer regions of the jets in 3C\,31, with initial conditions set as profiles of
emissivity and field-ordering across the jet at the start of the region. The
models fit reasonably well, but cannot accurately describe the observed
polarization at the jet edges.  A likely possibility is that the velocity field
is more complicated than the simple laminar flow we assume, and that some
turbulent component is present, leading to changes in field ordering and
strength which are not described by our model. In contrast, adiabatic models
fail completely in the inner and flaring regions. This should not come as a
surprise: the X-ray emission detected in these regions by \citet{Hard02} is most
likely to be synchrotron radiation, requiring significant particle acceleration.





\begin{thebibliography}{}



\bibitem[Baum et al.\ (1997)]{Bau97} Baum, S. A., et al. 1997, ApJ, 483, 178 (erratum ApJ, 492, 854)

\bibitem[Burch (1979)]{Bur79} Burch, S.F., 1979, MNRAS, 187, 187  

\bibitem[Hardcastle et al.\ (2002)]{Hard02} Hardcastle, M.J., Worrall,
 D.M., Birkinshaw, M., Laing, R.A., Bridle, A.H., 2002, MNRAS, 334, 182

\bibitem[Laing (1980)]{L80} Laing, R.A., 1980, MNRAS, 193, 439

\bibitem[Laing (2002)]{L02} Laing, R.A., 2002, MNRAS, 329, 417

\bibitem[Laing \& Bridle (2002a)]{LB02a} Laing, R.A., Bridle, A.H., 2002a,
MNRAS, 336, 328

\bibitem[Laing \& Bridle (2002b)]{LB02b} Laing, R.A., Bridle, A.H., 2002b,
MNRAS, 336, 1161

\bibitem[Laing et al.\ (1999)]{LPdRF} Laing, R. A., Parma, P., de Ruiter,
H. R., Fanti, R., 1999, MNRAS, 306, 513

\bibitem[Matthews \& Scheuer(1990)]{MS} Matthews, A.P., Scheuer, P.A.G., 1990, 
    MNRAS, 242, 616

\end{thebibliography}
\end{document}